\documentclass[%
 reprint,
 superscriptaddress,
nofootinbib,
 amsmath,amssymb,
 aps,
]{revtex4-2}

\usepackage{graphicx}
\usepackage{dcolumn}
\usepackage{bm}
\usepackage{array}
\newcolumntype{P}[1]{>{\centering\arraybackslash}p{#1}}
\usepackage[hidelinks]{hyperref}
\usepackage{booktabs}
\usepackage{upgreek}
\usepackage{comment}
\usepackage{verbatim}
\usepackage{amssymb}
\usepackage{multirow}
\usepackage{makecell}
\usepackage{aas_macros}

\usepackage[
separate-uncertainty = true,
multi-part-units=single
]{siunitx}
\usepackage{xcolor}
\usepackage[version=4]{mhchem}
\usepackage[normalem]{ulem}
\usepackage{tablefootnote}

\usepackage[section]{placeins} 

\usepackage{lineno}

\DeclareSIUnit\keVnr{keV_{nr}}
\DeclareSIUnit\keVee{keV_{ee}}

\newcommand\XeOneThreeSix{\ce{^{136}Xe}}

\newcommand\PoTwoOneFour{\ce{^{214}Po}}
\newcommand\BiTwoOneFour{\ce{^{214}Bi}}
\newcommand\TlTwoOEight{\ce{^{208}Tl}}

\newcommand\zvbb{0$\nu\beta\beta$}
\newcommand\us{$\mu$s}
\newcommand\um{$\mu$m}

\begin{document}


\title{Optical imaging in dual-phase xenon time projection chambers for neutrinoless double beta decay searches} 

\author{Adam D. Tidball}
\affiliation{Lawrence Livermore National Laboratory (LLNL), Livermore, CA 94550-9698, USA}    
\affiliation{University of California Davis, Department of Physics, One Shields Ave., Davis, CA 95616, USA}
\author{Jingke Xu}
\email[Corresponding Author: ]{xu12@llnl.gov}
\affiliation{Lawrence Livermore National Laboratory (LLNL), Livermore, CA 94550-9698, USA}    
\author{Brian G. Lenardo} \affiliation{SLAC National Accelerator Laboratory, Menlo Park, CA 94025, USA}  
\author{Nicholas Antipa} 
\affiliation{Department of Electrical and Computer Engineering, University of California San Diego, La Jolla, CA 92093-0112, USA}

\begin{abstract}
Dual-phase xenon time projection chambers (TPCs) are the leading technology for WIMP dark matter searches and have also demonstrated competitive sensitivities to neutrinoless double beta decays (\zvbb) of \XeOneThreeSix. However, the large energy disparity between these two signals requires different detector optimizations. In particular,  \zvbb\ searches in liquid xenon desire efficient rejection of closely located multiscatters of background gamma rays near the \zvbb\ energy, 
a capability achieved only in the vertical dimension in existing dual-phase xenon TPCs. 
We propose a highly customized, wide-angle optical system that can be conveniently integrated into large xenon TPCs and provide mm-scale transverse position resolution for MeV signals by imaging  gas electroluminescence light. This approach preserves a TPC's low-energy sensitivity and optimizes a single experiment for both dark matter and \zvbb\ searches. This paper discusses the design, simulated performance, and implications of the imaging system for future liquid xenon-based \zvbb\ searches. 
\end{abstract}

\maketitle

\emph{Introduction}---The observation of neutrinoless double beta decays (\zvbb) would provide direct evidence for lepton number violation and confirm the Majorana nature of neutrinos, shedding light on the origin of matter-antimatter asymmetry in the universe and neutrino mass hierarchy~\cite{neutrinolessDBD_prospect_status2019,Agostini2023_0vbb}.  
\XeOneThreeSix\ is a leading candidate for \zvbb\ decays, and xenon-based experiments have demonstrated exceptionally low backgrounds and high energy resolution, both of which are critical for the detection of \zvbb\ events. 
The use of gaseous and liquid xenon for \zvbb\ searches has been extensively investigated by experiments including NEXT-100~\cite{NEXT100_2026_detector}, EXO-200~\cite{EXO200_2019_fullexposure}, PandaX~\cite{PandaX2025_0vbb}, XENON1T~\cite{XENON2022_0vbb}, and LZ~\cite{LZ2021_0vbb}. Next-generation experiments such as NEXT~\cite{NEXT2021_0vbb}, nEXO~\cite{nEXO_sensitivity_2018} and XLZD~\cite{XLZD2025_0vbb} are expected to probe \zvbb\ half-life values in the range of 10$^{27}$--10$^{28}$ years. 

A \zvbb\ decay of \XeOneThreeSix\ is predicted to deposit 2.457~MeV energy, which, in the absence of antineutrino emission, is primarily shared by two electrons emitted from the decaying nucleus. For a well-shielded underground experiment, the dominant background in this energy region is high-energy gamma rays, especially those produced by decays of \BiTwoOneFour\ (2.447~MeV) and \TlTwoOEight\ (2.615~MeV) in detector components surrounding the active xenon target~\cite{PandaX2025_0vbb,XENON2022_0vbb,XLZD2025_0vbb,nEXO_sensitivity_2018}. Although \BiTwoOneFour\ may also be present inside xenon from radon emanation, this background can be promptly tagged by the immediate alpha decay of the daughter \PoTwoOneFour\ (t$_{1/2}$=164~\us). Rates of external gamma backgrounds are suppressed in the inner liquid xenon (LXe) volume but the large mean free path of gammas ($\sim$9~cm at 2.5~MeV) renders fiducialization inefficient due to the large loss of active target mass. 
Event topology, therefore, plays a critical role in signal and background discrimination. In LXe, $\sim$85\% of \zvbb\ events deposit all electron energy within a few millimeters, with the rest producing bremsstrahlung photons that travel further. Background gamma rays at 2.5~MeV, on the other hand, primarily Compton scatter, producing multiple vertices separated by several millimeters or larger. Fewer than 3\% of background gammas deposit all energy in a single interaction through photo-absorption. Therefore, a LXe detector that can separate ``single-site'' (SS) events from $\mathcal{O}$(mm)-separated ``multi-site'' (MS) events will efficiently distinguish \zvbb\ signals from MeV gamma backgrounds~\cite{Sazzad2026_SSMS}. 

In a dual-phase xenon time projection chamber (TPC), an event's vertical ($z$) position is estimated using the time required for ionization electrons to drift upwards to the detection region where electroluminescence (EL) light is produced. 
$\mathcal{O}$(mm) vertical position accuracy is achievable as the result of ionization pulse widths of $\sim$1~\us, electron drift velocities of $\sim$1~mm/\us, and longitudinal electron diffusion of a few mm~\cite{EXO2017_tDiffusion}. 
In the transverse ($x-y$) plane, however, event position information is only weakly encoded in the gradual change of detected EL light intensity in photosensors. Although mm-scale transverse position precision has been demonstrated for SS interactions~\cite{MIX2015,LUX2018_PositionReconstruction}, the TPC's ability to resolve transverse MS events such as gamma multi-scatters is significantly compromised by overlapping sensor patterns.  
Reference~\cite{Sazzad2026_SSMS} predicts that a TPC's SS/MS discrimination capability in the transverse plane could be improved if the individual sensor size, EL region thickness and sensor-EL region distance can be arbitrarily reduced. 
In practice, however, such configurations pose challenges to detector construction and background control; they may also not achieve the theoretical event reconstruction performance due to anticipated signal saturation in sensors. 

\emph{Method}---We address the transverse position reconstruction problem of dual-phase TPCs by investigating a customized imaging system that can be conveniently integrated into the TPC photosensor array and provide $\mathcal{O}$(mm) $x-y$ resolution for MeV signals. 
Dual-phase TPCs are designed to detect low-energy signals, and have demonstrated sensitivities to single-electron events with a typical EL amplification of $\sim$500~photons per extracted electron. A \zvbb\ event at $Q = 2.457$~MeV produces $\sim10^5$ ionized electrons, and would generate $\sim5\times10^7$ EL photons.
This large number of photons emitted from the highly localized EL region poses a challenge for existing TPC readouts but presents an opportunity for optical imaging techniques to be explored. 
The ARIADNE~\cite{ARIADNE2019_LArImaging} and NEXT-CRAB-0~\cite{CRAB2023_0vbbImaging} experiments have demonstrated $\sim$1~mm track reconstruction resolution in a liquid argon and a gaseous xenon TPC, respectively, by imaging EL light with conventional optics mounted outside the TPC with large standoff distances.
For a dual-phase xenon TPC, achieving low-energy dark matter sensitivity requires photosensors to be positioned near the EL region, impeding the use of imaging setups with small fields of view (FOV) from far above. 

\begin{figure}[t!]
\centering
\includegraphics[width=\linewidth]{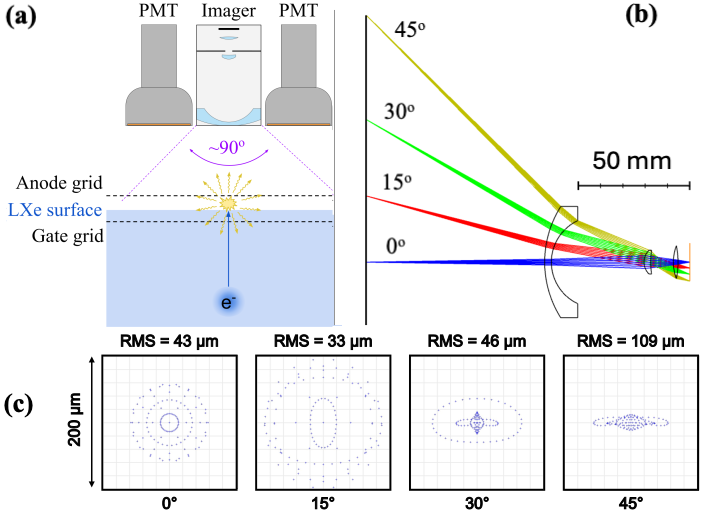}
\caption{{\bf a:} Illustration of optical imager deployment in a dual-phase TPC. {\bf b:} Lens design and illustrative VUV light paths for the customized wide-angle camera; only optical elements (lenses, stop, and sensor) are shown while supporting structures are not. The system has a diameter of 2 inches and a total length of $\sim$3 inches. {\bf c: } Spot diagrams for point sources at different positions on the object plane calculated by Zemax; at an effective magnification of 0.095, 200~\um\ in the image corresponds to $\sim$2~mm in the object plane. }
\label{fig1}
\end{figure}

We investigate a wide-angle camera design that employs a short focal length lens to image a large FOV using a package footprint on the scale of a PMT. This configuration can enable the imaging of a large area at a short distance using cameras embedded in the top photosensor array~\footnote{In principle, imaging from the bottom TPC photosensor array location is possible, but this design faces practical challenges from the requirement of enclosing alignment-sensitive optics in vacuum-sealed packages and significantly reduced photon fluxes.}, as illustrated in Fig.~\ref{fig1}a, allowing a full EL region coverage with a small fraction ($<$10\%) of TPC photosensors replaced with imagers, preserving the detector's light collection efficiency for dark matter searches. 

\emph{System Design}---Figure~\ref{fig1}b illustrates our proposed imager, designed to provide a high transmission of vacuum-UV (VUV) wavelengths and a $\lesssim$1~mm transverse position resolution within a 90$^\circ$ FOV (full field). At an 8~cm working distance, it is capable of imaging a circular area with a $\sim$8~cm radius, approximately 10 times larger than its own cross sectional area\footnote{We treat the FOV as circular, where the active sensor area circumscribes the lens's image circle}. 
The optics consists of 3 lenses; the large front negative lens, assumed to be VUV-grade synthetic fused silica for its low radioactivity level and relatively high VUV transmission, gathers light from a $\sim$90$^\circ$ FOV, 
and two smaller positive lenses in the rear, modeled as CaF$_2$ for high VUV transparency, correct for aberrations introduced by the large incident photon angle range and create in-focus images. An optical stop is placed in between the second and third lenses to block stray light rays for image quality. At 175~nm, ray tracing simulations of the uncoated lenses using manufacturer material data give an on-axis transmittance of about 73\%, with losses dominated by Fresnel reflection ($\sim$24\%; 5.3\% per fused silica surface and 4.3\% per CaF$_2$ surface) rather than bulk absorption ($\sim$3\%, almost entirely in the fused silica).

This design is inspired by commercial wide angle lenses that achieve up to $\gtrsim$180$^\circ$ FOV with around ten optical elements, 
but is customized to reduce the total number of elements to mitigate transmission loss of 175~nm xenon EL light and radioactive backgrounds from the components. 
The lens assembly is designed to operate at 175 nm wavelength with an image-space f-number of 2.8, angular full FOV of $90^\circ$, effective focal length of 10 mm, and paraxial magnification of 0.095. 
Such performance goals typically require many elements to control aberrations, but our simplified design is made possible by the modest position resolution requirement of $\gtrsim$1~mm at the object. This corresponds to an angular imaging resolution approximately 10 times larger than that of typical wide-angle lenses. 
Higher position resolution may be achieved through a more complex lens design (e.g., by introducing aspheres or additional lens elements), but this is not expected to significantly improve the gamma background rejection power due to the transverse diffusion, vertical span of the EL region, and effects from grid focusing, all of which are at the  millimeter scale and will be discussed in the following sections. 

The optical performance of the imager is characterized with Zemax OpticStudio, which was also used for design optimization. 
Figure~\ref{fig1}c shows the calculated sensor-space spot diagrams for point sources at different fields of view. 
As the field angle increases, the point-spread function (PSF) becomes asymmetric due to off-axis aberrations, but the root mean square (RMS) errors remain $\lesssim$1~mm in the object space.
The light collection efficiency of the imager, including transmission loss, is estimated using nonsequential Monte Carlo simulations of OpticsStudio. We find an efficiency of 5$\times10^{-5}$ for sources in the center of the FOV, which gradually decreases to 3$\times10^{-5}$ for sources at 45$^\circ$, leading to $\sim$2000 photons delivered to the imaging sensor for an event in the \zvbb\ energy region. 
The sensor is assumed to have a size of a 1.6~cm$\times$1.6~cm with 100~\um\ pixel size. 
In this study, we use an image exposure time of 3~ms, slightly larger than the maximum electron drift time expected of an XLZD-sized TPC. 
No sub-exposure timing sensitivity is required for the camera sensor, but a time resolved readout such as that demonstrated in \cite{ARIADNE2019_LArImaging,Fischer2025_dsipm}, if implemented, can enable 3D imaging with improved transverse position resolutions. 

\emph{Simulated Images}---To study the imager's performance in differentiating \zvbb\ events and gamma backgrounds at the same energy, we simulate the energy depositions of such interactions in LXe with Geant4, convolve the ionization signals with a Gaussian transverse diffusion kernel, and create spatially diffused light sources in the EL region. 
For each simulated EL signal, a total of 5$\times10^7$ photons at 175~nm wavelength are emitted isotropically from a vertical span of 5~mm, which represents the typical EL gap used in xenon TPCs, as in the LZ experiment~\cite{linehanDesignProductionHigh2022}. 
These photons are then transported through optics to form images on the sensor to be used in the following analysis. 
The low light collection efficiency is the dominant factor in observed photon statistics, so fluctuations in the emitted photon count is neglected. 

\begin{figure}[t!]
\centering
\includegraphics[width=\linewidth]{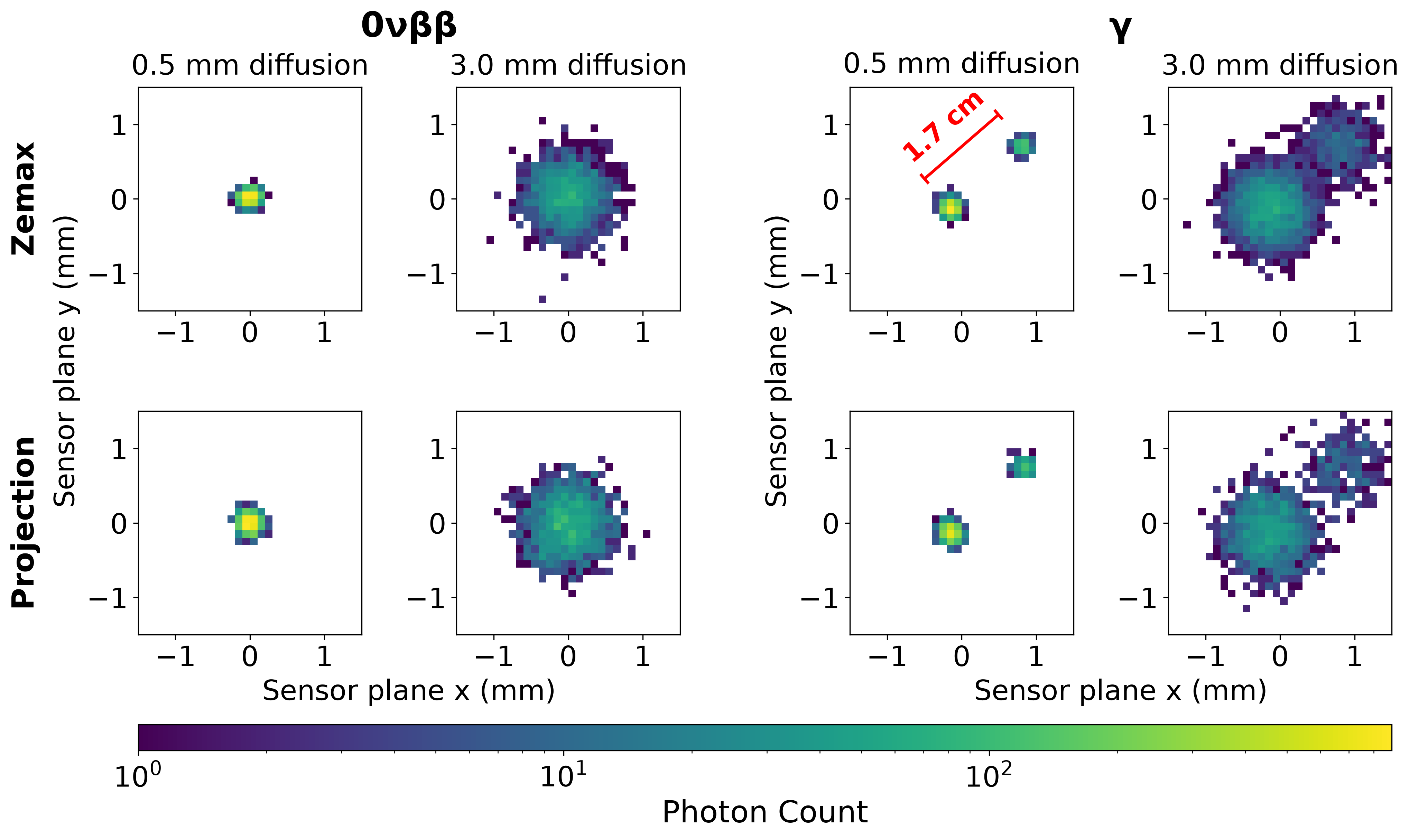}
\caption{Simulated sensor images for example \zvbb\ (SS) and gamma (MS) interactions at the center of the object plane with different transverse electron diffusion assumptions. Coordinates are given in the sensor plane, which carries a de-magnification factor of 0.095 relative to the object space; the physical separation of the two gamma interaction vertices in object space is $\sim$1.7 cm. The top row images are computed using the Zemax PSF and the bottom row ones are approximated with a fast projection algorithm using a Gaussian smearing. In this example, the sensor is assumed to be single-photon-sensitive with 100\% quantum efficiency (QE).}
\label{fig2}
\end{figure} 

Figure 2 (top) shows Zemax-computed images for a SS \zvbb\ event and a double-scatter (DS) gamma interaction at the center of the FOV assuming 0.5~mm and 3~mm transverse diffusion, respectively. 
The 3~mm value represents the average diffusion in a 3~m tall XLZD-sized xenon TPC, assuming a 1~ms drift time and a 55~cm$^2$/s transverse diffusion coefficient at $\sim$100~V/cm drift field~\cite{EXO2017_tDiffusion}; the 0.5~mm study illustrates the system's performance with minimal transverse diffusion, such as events within the top 10~cm of the TPC or in a TPC with a diffusion-suppression dopant added to LXe, as has been demonstrated in gaseous TPCs~\cite{NEXT2019_diffusion}. 
The gamma ray event shown in Figure~\ref{fig2} illustrates the most challenging background pathology in \zvbb\ searches: the first gamma interaction deposits nearly 90\% of its energy in a 180$^\circ$ Compton scatter, and then the scattered low-energy gamma ray is photoabsorbed $\sim$1~cm away in the same transverse plane. 
Such MS events in a dual-phase xenon TPC cannot be identified using drift time, and the two closely located vertices, combined with their large energy disparity, makes them extremely challenging to resolve using conventional readout~\cite{Sazzad2026_SSMS}. 
Our proposed optical imager could clearly reveal the vertices in the low-diffusion scenario, and allow statistical signal-background discrimination with modest diffusion.

To systematically evaluate the signal-background discrimination capability of the imager with high statistic simulations, we have developed a fast image generation algorithm that projects point sources from the object space onto a sensor pixel using the Zemax-predicted field distortion. Instead of running simulations in Zemax with the exact PSF structures illustrated in Fig.~\ref{fig1}c, we use a simple Gaussian smearing based on the Zemax-calculated RMS error for each PSF.  
The 175~nm photon collection efficiency estimated by Zemax is also parameterized as a function of source location, which introduces  binominal fluctuations in the detected photon count. 
The axial symmetry of the optical model and the narrow depth of field allow the projection to be carried out using a single parameter, which is the field angle. 
This approach greatly speeds up the image simulations and only introduces minimal inaccuracies even for tracks with low diffusion (0.5~mm), as illustrated in Fig. 2 (bottom). In addition, images simulated using this approach are on average more smeared than those predicted by Zemax, rendering the analysis results conservative. 

\emph{\zvbb\ sensitivity}---We simulate 10,000 each of \zvbb\ events and 2.5~MeV gamma backgrounds in Geant4 and create images using the projection method. The \zvbb\ simulation assumes the standard light neutrino propagator and generates the two electrons with energies and angular distributions based on the allowed phase space~\cite{Boehm_Vogel_1992}. Because gamma multi-scatter interactions in the $z$ dimension can already be well-resolved using the standard TPC readout, we restrict our analysis to events with $>$97\% energy deposited within a 5~mm depth slice of LXe. 
This selection criterion rejects 89.5\% \BiTwoOneFour\ gamma backgrounds and preserves 84.5\% \zvbb\ signals, and is tuned to match the expected signal/background efficiencies for a large TPC with 3~mm MS resolving power in the $z$ dimension reported in earlier studies~\cite{Darwin2020_0vbb,XLZD2025_0vbb}. 
For each simulated image, we use a photon-weighted DBSCAN algorithm \cite{ester1996density,sander1998density} to determine the multiplicity of interaction vertices in the FOV, and reject events with more than one cluster as gamma background candidates. We optimize the DBSCAN parameters (distance between activated pixels and minimum samples in a cluster) for each choice of electron diffusion, sensor pixel size, and sensor QE. 
For images with single identified clusters, we use skewness of the cluster along the principal component direction to statistically reject merged MS gamma backgrounds, which are typically less symmetric than SS \zvbb\ signals, as illustrated in Fig.~\ref{fig2}. 

\begin{figure}[t!]
\centering
\includegraphics[width=\linewidth]{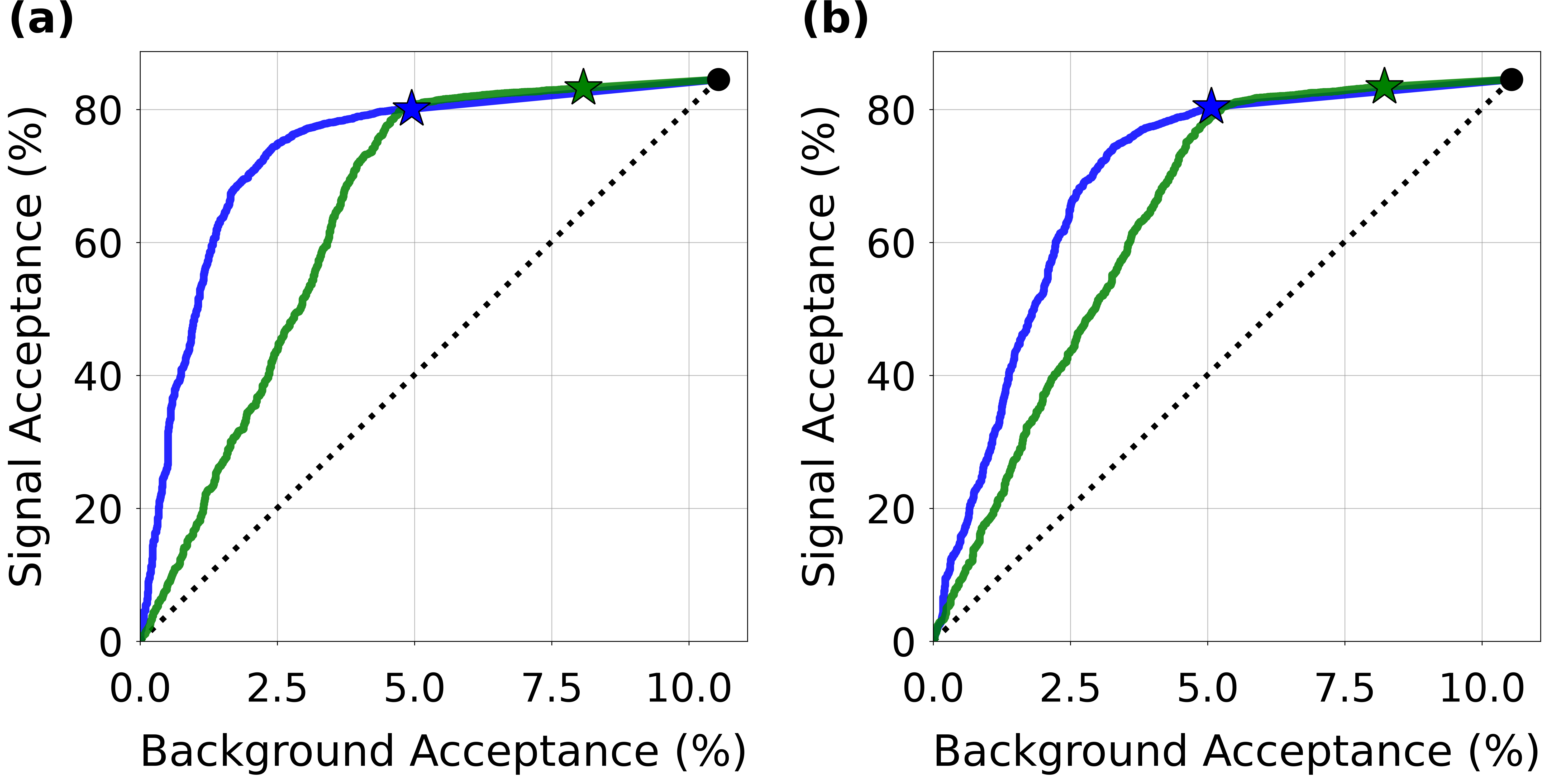}
\caption{Evaluated \zvbb\ signal acceptance and 2.5~MeV gamma leakage fractions for events occurring right below an imager for 0.5~mm (blue) and 3~mm (green) transverse diffusion. {\bf (a)} assumes a linear sensor with 50\% detection efficiency for 175~nm photons with a typical sCMOS read noise of 1~electron, and {\bf (b)} assumes a binary SPAD array with 20\% photon detection efficiency, 0.19~Hz/mm$^2$ dark count rate~\cite{nEXO2022_SiPM}, and 3~ms exposure time. Both sensor models use a pixel size of 100~\um. The straight lines connecting the ``dot'' and ``star'' symbols in the upper right corners represent the background reduction effect of a single-cluster cut, and the curved lines to the left correspond to a skewness cut for single-cluster images. $\sim$3\% of all 2.5~MeV gamma interactions may deposit all energies at a single vertex through photoabsorption. }
\label{fig3}
\end{figure}

Figure 3 shows the predicted \zvbb\ signal acceptance as a function of gamma background leakage for simulated events in the center of the FOV, using the single-cluster cut and skewness cut, where (a) assumes a linear sCMOS sensor and (b) assumes a binary SPAD array sensor. 
For both sensor models, when a 0.5~mm transverse electron diffusion is applied, our lens design preserves sharp vertex features, allowing $\sim$1/2 of the residual gamma backgrounds in conventional dual-phase xenon TPCs to be rejected with the single-cluster cut alone; additional background rejection using the skewness cut enables nearly all MS gamma backgrounds to be identified at a $\sim$5\% loss of \zvbb\ signals. When a 3~mm diffusion is assumed, the images become blurry, causing nearby MS vertices to merge and the single-cluster cut to lose efficiency. However, when both cuts are combined, this simple analysis still rejects $\gtrsim$1/2 of the gamma backgrounds with little loss of \zvbb\ signals.  
For events at non-zero field angles, the $z$-depth of the EL tracks becomes visible to the imager and contribute additional blurring to the vertex images; this causes a small degradation of signal-background discrimination power for events with 0.5~mm diffusion, and its effect is less noticeable for already-smeared events with 3~mm diffusion. 
It should also be noted that this image distortion is determined by the field angle, and, in principle, can be accurately modeled to further mitigate the efficiency loss. 
In addition, our imager is designed to reconstruct point sources between -45$^\circ$ to +45$^\circ$ field angles with $\lesssim$1~mm RMS errors, but can be further optimized to also record events occurring at larger radii with modestly increased spot sizes; so events at large field angles for one imager can be observed by additional imagers, allowing computational stereoimaging that further compensates losses of reconstruction accuracy for these events. 

In this work, we assume that the TPC perfectly preserves the $x-y$ structure of the initial ionization event in LXe. In reality, however, a TPC distorts the ionization track topology due to imperfect drift and extraction fields. 
Most importantly, the geometry of the TPC extraction grid can create a strong electron focusing effect at the scale of the grid pitch, causing a continuous electron cloud to become semi-discrete near the liquid surface~\cite{MIX2015,LUX2018_PositionReconstruction}. On the one hand, this grid focusing effect creates sharper EL features that increase the signal-to-noise ratio of the images; on the other hand, it can produce multiple vertices in SS event images, preventing the use of our simple cluster-based analysis. 
The implications of this effect are conservatively evaluated by increasing the sensor pixel size in our analysis from 100~\um\ to 500~\um, which effectively blurs the predicted images and homogenizes sharp image features introduced by grid focusing; the 500~\um\ value corresponds to a pixel resolution of $\sim$5~mm in the object space, which is the typical grid pitch used in xenon TPCs. This study suggests that a modest performance degradation could occur for events with 0.5~mm diffusion but events with 3~mm diffusion are minimally affected. 
It is worth noting that this grid focusing effect can be accurately addressed with an analysis that employs pattern-recognition classification algorithms for images acquired with smaller pixels than the grid pitch. 
The imager design, the pattern-recognition analysis, and the TPC grid pitch can be co-optimized to maximally benefit from the signal enhancement effect while mitigating biases from electron focusing. 

\emph{Conclusion}---This work investigates a highly customized optical imager that can be conveniently integrated in dual-phase xenon TPCs to resolve fine features in the transverse plane. We demonstrate through simulations that our imager design can suppress residual 2.5~MeV gamma backgrounds in conventional dual-phase xenon TPCs by a factor of 2--3, increasing the experiments' \zvbb\ sensitivities. 
Importantly, this imaging approach is compatible with the current design of the XLZD detector and only requires $<$10\% of the gas-phase photosensors to be replaced with low-radioactivity, PMT-shaped imagers. This modest detector upgrade can substantially improve the detector's \zvbb\ sensitivity while preserving its energy and position performance for low-energy dark matter searches, 
allowing both high profile science goals to be pursued simultaneously. 

The system's performance can be further improved in future work. By exploring additional design space, the optical imager can be optimized for larger FOV values, allowing 3D image reconstruction from imagers at different view angles, which can also be achieved by incorporating sub-microsecond-speed sensors. 
Future analyses will model and correct for image distortions from large field angles and grid focusing, while incorporating the imager's calculated PSF instead of a simplified 2D Gaussian smearing and considering higher order optical effects, such as light reflection off detector surfaces and shadowing of EL light by TPC grids. 
We will also correlate imaging data with TPC pulse shape information to explore reconstruction improvements based on complementarity between this technique's unique sensitivity to transverse MS events and the TPC's sensitivity to vertical MS vertices.

\emph{Acknowledgments}---This work was performed under the auspices of the U.S. Department of Energy (DOE) by Lawrence Livermore National Laboratory (LLNL) under Contract DE-AC52-07NA27344. 
It is supported by the LLNL LDRD program under award 25-LW-034. A.T. was partially supported by the DOE/NNSA under Award Number DE-NA0000979 through the Nuclear Science and Security Consortium.  
LLNL IM release number: LLNL-JRNL-2021151.

\bibliographystyle{apsrev4-2}
\bibliography{main}

\begin{thebibliography}{24}%
\makeatletter
\providecommand \@ifxundefined [1]{%
 \@ifx{#1\undefined}
}%
\providecommand \@ifnum [1]{%
 \ifnum #1\expandafter \@firstoftwo
 \else \expandafter \@secondoftwo
 \fi
}%
\providecommand \@ifx [1]{%
 \ifx #1\expandafter \@firstoftwo
 \else \expandafter \@secondoftwo
 \fi
}%
\providecommand \natexlab [1]{#1}%
\providecommand \enquote  [1]{``#1''}%
\providecommand \bibnamefont  [1]{#1}%
\providecommand \bibfnamefont [1]{#1}%
\providecommand \citenamefont [1]{#1}%
\providecommand \href@noop [0]{\@secondoftwo}%
\providecommand \href [0]{\begingroup \@sanitize@url \@href}%
\providecommand \@href[1]{\@@startlink{#1}\@@href}%
\providecommand \@@href[1]{\endgroup#1\@@endlink}%
\providecommand \@sanitize@url [0]{\catcode `\\12\catcode `\$12\catcode
  `\&12\catcode `\#12\catcode `\^12\catcode `\_12\catcode `\%12\relax}%
\providecommand \@@startlink[1]{}%
\providecommand \@@endlink[0]{}%
\providecommand \url  [0]{\begingroup\@sanitize@url \@url }%
\providecommand \@url [1]{\endgroup\@href {#1}{\urlprefix }}%
\providecommand \urlprefix  [0]{URL }%
\providecommand \Eprint [0]{\href }%
\providecommand \doibase [0]{https://doi.org/}%
\providecommand \selectlanguage [0]{\@gobble}%
\providecommand \bibinfo  [0]{\@secondoftwo}%
\providecommand \bibfield  [0]{\@secondoftwo}%
\providecommand \translation [1]{[#1]}%
\providecommand \BibitemOpen [0]{}%
\providecommand \bibitemStop [0]{}%
\providecommand \bibitemNoStop [0]{.\EOS\space}%
\providecommand \EOS [0]{\spacefactor3000\relax}%
\providecommand \BibitemShut  [1]{\csname bibitem#1\endcsname}%
\let\auto@bib@innerbib\@empty
\bibitem [{\citenamefont {Dolinski}\ \emph {et~al.}(2019)\citenamefont
  {Dolinski}, \citenamefont {Poon},\ and\ \citenamefont
  {Rodejohann}}]{neutrinolessDBD_prospect_status2019}%
  \BibitemOpen
  \bibfield  {author} {\bibinfo {author} {\bibfnamefont {M.~J.}\ \bibnamefont
  {Dolinski}}, \bibinfo {author} {\bibfnamefont {A.~W.}\ \bibnamefont {Poon}},\
  and\ \bibinfo {author} {\bibfnamefont {W.}~\bibnamefont {Rodejohann}},\
  }\href {https://doi.org/https://doi.org/10.1146/annurev-nucl-101918-023407}
  {\bibfield  {journal} {\bibinfo  {journal} {Annual Review of Nuclear and
  Particle Science}\ }\textbf {\bibinfo {volume} {69}},\ \bibinfo {pages} {219}
  (\bibinfo {year} {2019})}\BibitemShut {NoStop}%
\bibitem [{\citenamefont {Agostini}\ \emph {et~al.}(2023)\citenamefont
  {Agostini}, \citenamefont {Benato}, \citenamefont {Detwiler}, \citenamefont
  {Men\'endez},\ and\ \citenamefont {Vissani}}]{Agostini2023_0vbb}%
  \BibitemOpen
  \bibfield  {author} {\bibinfo {author} {\bibfnamefont {M.}~\bibnamefont
  {Agostini}}, \bibinfo {author} {\bibfnamefont {G.}~\bibnamefont {Benato}},
  \bibinfo {author} {\bibfnamefont {J.~A.}\ \bibnamefont {Detwiler}}, \bibinfo
  {author} {\bibfnamefont {J.}~\bibnamefont {Men\'endez}},\ and\ \bibinfo
  {author} {\bibfnamefont {F.}~\bibnamefont {Vissani}},\ }\href
  {https://doi.org/10.1103/RevModPhys.95.025002} {\bibfield  {journal}
  {\bibinfo  {journal} {Rev. Mod. Phys.}\ }\textbf {\bibinfo {volume} {95}},\
  \bibinfo {pages} {025002} (\bibinfo {year} {2023})}\BibitemShut {NoStop}%
\bibitem [{\citenamefont {Adams}\ \emph {et~al.}(2026)\citenamefont {Adams}
  \emph {et~al.}}]{NEXT100_2026_detector}%
  \BibitemOpen
  \bibfield  {author} {\bibinfo {author} {\bibfnamefont {C.}~\bibnamefont
  {Adams}} \emph {et~al.} (\bibinfo {collaboration} {NEXT Collaboration}),\
  }\href {https://doi.org/10.1140/epjc/s10052-025-14951-y} {\bibfield
  {journal} {\bibinfo  {journal} {The European Physical Journal C}\ }\textbf
  {\bibinfo {volume} {86}},\ \bibinfo {pages} {114} (\bibinfo {year}
  {2026})}\BibitemShut {NoStop}%
\bibitem [{\citenamefont {Anton}\ \emph {et~al.}(2019)\citenamefont {Anton}
  \emph {et~al.}}]{EXO200_2019_fullexposure}%
  \BibitemOpen
  \bibfield  {author} {\bibinfo {author} {\bibfnamefont {G.}~\bibnamefont
  {Anton}} \emph {et~al.} (\bibinfo {collaboration} {EXO-200 Collaboration}),\
  }\href {https://doi.org/10.1103/PhysRevLett.123.161802} {\bibfield  {journal}
  {\bibinfo  {journal} {Phys. Rev. Lett.}\ }\textbf {\bibinfo {volume} {123}},\
  \bibinfo {pages} {161802} (\bibinfo {year} {2019})}\BibitemShut {NoStop}%
\bibitem [{\citenamefont {Zhang}\ \emph {et~al.}(2025)\citenamefont {Zhang}
  \emph {et~al.}}]{PandaX2025_0vbb}%
  \BibitemOpen
  \bibfield  {author} {\bibinfo {author} {\bibfnamefont {S.}~\bibnamefont
  {Zhang}} \emph {et~al.} (\bibinfo {collaboration} {PandaX Collaboration}),\
  }\href {https://doi.org/https://doi.org/10.1016/j.scib.2025.03.009}
  {\bibfield  {journal} {\bibinfo  {journal} {Science Bulletin}\ }\textbf
  {\bibinfo {volume} {70}},\ \bibinfo {pages} {1779} (\bibinfo {year}
  {2025})}\BibitemShut {NoStop}%
\bibitem [{\citenamefont {Aprile}\ \emph {et~al.}(2022)\citenamefont {Aprile}
  \emph {et~al.}}]{XENON2022_0vbb}%
  \BibitemOpen
  \bibfield  {author} {\bibinfo {author} {\bibfnamefont {E.}~\bibnamefont
  {Aprile}} \emph {et~al.} (\bibinfo {collaboration} {XENON Collaboration}),\
  }\href {https://doi.org/10.1103/PhysRevC.106.024328} {\bibfield  {journal}
  {\bibinfo  {journal} {Phys. Rev. C}\ }\textbf {\bibinfo {volume} {106}},\
  \bibinfo {pages} {024328} (\bibinfo {year} {2022})}\BibitemShut {NoStop}%
\bibitem [{\citenamefont {Akerib}\ \emph {et~al.}(2021)\citenamefont {Akerib}
  \emph {et~al.}}]{LZ2021_0vbb}%
  \BibitemOpen
  \bibfield  {author} {\bibinfo {author} {\bibfnamefont {D.~S.}\ \bibnamefont
  {Akerib}} \emph {et~al.} (\bibinfo {collaboration} {The LUX-ZEPLIN
  Collaboration}),\ }\href {https://doi.org/10.1103/PhysRevC.104.065501}
  {\bibfield  {journal} {\bibinfo  {journal} {Phys. Rev. C}\ }\textbf {\bibinfo
  {volume} {104}},\ \bibinfo {pages} {065501} (\bibinfo {year}
  {2021})}\BibitemShut {NoStop}%
\bibitem [{\citenamefont {Adams}\ \emph {et~al.}(2021)\citenamefont {Adams}
  \emph {et~al.}}]{NEXT2021_0vbb}%
  \BibitemOpen
  \bibfield  {author} {\bibinfo {author} {\bibfnamefont {C.}~\bibnamefont
  {Adams}} \emph {et~al.} (\bibinfo {collaboration} {NEXT Collaboration}),\
  }\href {https://doi.org/10.1007/JHEP08(2021)164} {\bibfield  {journal}
  {\bibinfo  {journal} {Journal of High Energy Physics}\ }\textbf {\bibinfo
  {volume} {2021}},\ \bibinfo {pages} {164} (\bibinfo {year}
  {2021})}\BibitemShut {NoStop}%
\bibitem [{\citenamefont {Albert}\ \emph {et~al.}(2018)\citenamefont {Albert}
  \emph {et~al.}}]{nEXO_sensitivity_2018}%
  \BibitemOpen
  \bibfield  {author} {\bibinfo {author} {\bibfnamefont {J.~B.}\ \bibnamefont
  {Albert}} \emph {et~al.} (\bibinfo {collaboration} {nEXO Collaboration}),\
  }\href {https://doi.org/10.1103/PhysRevC.97.065503} {\bibfield  {journal}
  {\bibinfo  {journal} {Phys. Rev. C}\ }\textbf {\bibinfo {volume} {97}},\
  \bibinfo {pages} {065503} (\bibinfo {year} {2018})}\BibitemShut {NoStop}%
\bibitem [{\citenamefont {Aalbers}\ \emph {et~al.}(2025)\citenamefont {Aalbers}
  \emph {et~al.}}]{XLZD2025_0vbb}%
  \BibitemOpen
  \bibfield  {author} {\bibinfo {author} {\bibfnamefont {J.}~\bibnamefont
  {Aalbers}} \emph {et~al.} (\bibinfo {collaboration} {XLZD Collaboration}),\
  }\href {https://doi.org/10.1088/1361-6471/adb900} {\bibfield  {journal}
  {\bibinfo  {journal} {Journal of Physics G: Nuclear and Particle Physics}\
  }\textbf {\bibinfo {volume} {52}},\ \bibinfo {pages} {045102} (\bibinfo
  {year} {2025})}\BibitemShut {NoStop}%
\bibitem [{\citenamefont {Sazzad}\ \emph {et~al.}(2026)\citenamefont {Sazzad},
  \citenamefont {Hardy}, \citenamefont {Dai}, \citenamefont {Xu}, \citenamefont
  {Lenardo}, \citenamefont {Sutanto}, \citenamefont {Antipa}, \citenamefont
  {Koertzen}, \citenamefont {John}, \citenamefont {Akinin},\ and\ \citenamefont
  {Pershing}}]{Sazzad2026_SSMS}%
  \BibitemOpen
  \bibfield  {author} {\bibinfo {author} {\bibfnamefont {A.}~\bibnamefont
  {Sazzad}}, \bibinfo {author} {\bibfnamefont {C.}~\bibnamefont {Hardy}},
  \bibinfo {author} {\bibfnamefont {X.}~\bibnamefont {Dai}}, \bibinfo {author}
  {\bibfnamefont {J.}~\bibnamefont {Xu}}, \bibinfo {author} {\bibfnamefont
  {B.}~\bibnamefont {Lenardo}}, \bibinfo {author} {\bibfnamefont
  {F.}~\bibnamefont {Sutanto}}, \bibinfo {author} {\bibfnamefont
  {N.}~\bibnamefont {Antipa}}, \bibinfo {author} {\bibfnamefont
  {J.}~\bibnamefont {Koertzen}}, \bibinfo {author} {\bibfnamefont
  {P.}~\bibnamefont {John}}, \bibinfo {author} {\bibfnamefont {A.}~\bibnamefont
  {Akinin}},\ and\ \bibinfo {author} {\bibfnamefont {T.}~\bibnamefont
  {Pershing}},\ }\href {https://doi.org/10.1088/1748-0221/21/02/P02016}
  {\bibfield  {journal} {\bibinfo  {journal} {Journal of Instrumentation}\
  }\textbf {\bibinfo {volume} {21}}\bibinfo  {number} { (02)},\ \bibinfo
  {pages} {P02016}}\BibitemShut {NoStop}%
\bibitem [{\citenamefont {Albert}\ \emph {et~al.}(2017)\citenamefont {Albert}
  \emph {et~al.}}]{EXO2017_tDiffusion}%
  \BibitemOpen
\bibfield  {number} {  }\bibfield  {author} {\bibinfo {author} {\bibfnamefont
  {J.~B.}\ \bibnamefont {Albert}} \emph {et~al.} (\bibinfo {collaboration}
  {EXO-200 Collaboration}),\ }\href
  {https://doi.org/10.1103/PhysRevC.95.025502} {\bibfield  {journal} {\bibinfo
  {journal} {Phys. Rev. C}\ }\textbf {\bibinfo {volume} {95}},\ \bibinfo
  {pages} {025502} (\bibinfo {year} {2017})}\BibitemShut {NoStop}%
\bibitem [{\citenamefont {Stephenson}\ \emph {et~al.}(2015)\citenamefont
  {Stephenson}, \citenamefont {Haefner}, \citenamefont {Lin}, \citenamefont
  {Ni}, \citenamefont {Pushkin}, \citenamefont {Raymond}, \citenamefont
  {Schubnell}, \citenamefont {Shutty}, \citenamefont {Tarlé}, \citenamefont
  {Weaverdyck},\ and\ \citenamefont {Lorenzon}}]{MIX2015}%
  \BibitemOpen
  \bibfield  {author} {\bibinfo {author} {\bibfnamefont {S.}~\bibnamefont
  {Stephenson}}, \bibinfo {author} {\bibfnamefont {J.}~\bibnamefont {Haefner}},
  \bibinfo {author} {\bibfnamefont {Q.}~\bibnamefont {Lin}}, \bibinfo {author}
  {\bibfnamefont {K.}~\bibnamefont {Ni}}, \bibinfo {author} {\bibfnamefont
  {K.}~\bibnamefont {Pushkin}}, \bibinfo {author} {\bibfnamefont
  {R.}~\bibnamefont {Raymond}}, \bibinfo {author} {\bibfnamefont
  {M.}~\bibnamefont {Schubnell}}, \bibinfo {author} {\bibfnamefont
  {N.}~\bibnamefont {Shutty}}, \bibinfo {author} {\bibfnamefont
  {G.}~\bibnamefont {Tarlé}}, \bibinfo {author} {\bibfnamefont
  {C.}~\bibnamefont {Weaverdyck}},\ and\ \bibinfo {author} {\bibfnamefont
  {W.}~\bibnamefont {Lorenzon}},\ }\href
  {https://doi.org/10.1088/1748-0221/10/10/P10040} {\bibfield  {journal}
  {\bibinfo  {journal} {Journal of Instrumentation}\ }\textbf {\bibinfo
  {volume} {10}}\bibinfo  {number} { (10)},\ \bibinfo {pages}
  {P10040}}\BibitemShut {NoStop}%
\bibitem [{\citenamefont {Akerib}\ \emph {et~al.}(2018)\citenamefont {Akerib},
  \citenamefont {Alsum}, \citenamefont {Araújo}, \citenamefont {Bai} \emph
  {et~al.}}]{LUX2018_PositionReconstruction}%
  \BibitemOpen
\bibfield  {number} {  }\bibfield  {author} {\bibinfo {author} {\bibfnamefont
  {D.}~\bibnamefont {Akerib}}, \bibinfo {author} {\bibfnamefont
  {S.}~\bibnamefont {Alsum}}, \bibinfo {author} {\bibfnamefont
  {H.}~\bibnamefont {Araújo}}, \bibinfo {author} {\bibfnamefont
  {X.}~\bibnamefont {Bai}}, \emph {et~al.},\ }\href
  {https://doi.org/10.1088/1748-0221/13/02/P02001} {\bibfield  {journal}
  {\bibinfo  {journal} {Journal of Instrumentation}\ }\textbf {\bibinfo
  {volume} {13}}\bibinfo  {number} { (02)},\ \bibinfo {pages}
  {P02001}}\BibitemShut {NoStop}%
\bibitem [{\citenamefont {Roberts}\ \emph {et~al.}(2019)\citenamefont
  {Roberts}, \citenamefont {Svihra}, \citenamefont {Al-Refaie}, \citenamefont
  {Graafsma}, \citenamefont {Küpper}, \citenamefont {Majumdar}, \citenamefont
  {Mavrokoridis}, \citenamefont {Nomerotski}, \citenamefont {Pennicard},
  \citenamefont {Philippou}, \citenamefont {Trippel}, \citenamefont
  {Touramanis},\ and\ \citenamefont {Vann}}]{ARIADNE2019_LArImaging}%
  \BibitemOpen
\bibfield  {number} {  }\bibfield  {author} {\bibinfo {author} {\bibfnamefont
  {A.}~\bibnamefont {Roberts}}, \bibinfo {author} {\bibfnamefont
  {P.}~\bibnamefont {Svihra}}, \bibinfo {author} {\bibfnamefont
  {A.}~\bibnamefont {Al-Refaie}}, \bibinfo {author} {\bibfnamefont
  {H.}~\bibnamefont {Graafsma}}, \bibinfo {author} {\bibfnamefont
  {J.}~\bibnamefont {Küpper}}, \bibinfo {author} {\bibfnamefont
  {K.}~\bibnamefont {Majumdar}}, \bibinfo {author} {\bibfnamefont
  {K.}~\bibnamefont {Mavrokoridis}}, \bibinfo {author} {\bibfnamefont
  {A.}~\bibnamefont {Nomerotski}}, \bibinfo {author} {\bibfnamefont
  {D.}~\bibnamefont {Pennicard}}, \bibinfo {author} {\bibfnamefont
  {B.}~\bibnamefont {Philippou}}, \bibinfo {author} {\bibfnamefont
  {S.}~\bibnamefont {Trippel}}, \bibinfo {author} {\bibfnamefont
  {C.}~\bibnamefont {Touramanis}},\ and\ \bibinfo {author} {\bibfnamefont
  {J.}~\bibnamefont {Vann}},\ }\href
  {https://doi.org/10.1088/1748-0221/14/06/P06001} {\bibfield  {journal}
  {\bibinfo  {journal} {Journal of Instrumentation}\ }\textbf {\bibinfo
  {volume} {14}}\bibinfo  {number} { (06)},\ \bibinfo {pages}
  {P06001}}\BibitemShut {NoStop}%
\bibitem [{\citenamefont {Byrnes}\ \emph {et~al.}(2023)\citenamefont {Byrnes}
  \emph {et~al.}}]{CRAB2023_0vbbImaging}%
  \BibitemOpen
\bibfield  {number} {  }\bibfield  {author} {\bibinfo {author} {\bibfnamefont
  {N.}~\bibnamefont {Byrnes}} \emph {et~al.} (\bibinfo {collaboration} {NEXT
  Collaboration}),\ }\href {https://doi.org/10.1088/1748-0221/18/08/P08006}
  {\bibfield  {journal} {\bibinfo  {journal} {Journal of Instrumentation}\
  }\textbf {\bibinfo {volume} {18}}\bibinfo  {number} { (08)},\ \bibinfo
  {pages} {P08006}}\BibitemShut {NoStop}%
\bibitem [{\citenamefont {Fischer}\ \emph {et~al.}(2025)\citenamefont {Fischer}
  \emph {et~al.}}]{Fischer2025_dsipm}%
  \BibitemOpen
\bibfield  {number} {  }\bibfield  {author} {\bibinfo {author} {\bibfnamefont
  {P.}~\bibnamefont {Fischer}} \emph {et~al.},\ }\href
  {https://doi.org/10.1088/1748-0221/20/06/C06008} {\bibfield  {journal}
  {\bibinfo  {journal} {J. Instrum.}\ }\textbf {\bibinfo {volume} {20}}\bibinfo
   {number} { (06)},\ \bibinfo {pages} {C06008}}\BibitemShut {NoStop}%
\bibitem [{\citenamefont {Linehan}\ \emph {et~al.}(2022)\citenamefont {Linehan}
  \emph {et~al.}}]{linehanDesignProductionHigh2022}%
  \BibitemOpen
\bibfield  {number} {  }\bibfield  {author} {\bibinfo {author} {\bibfnamefont
  {R.}~\bibnamefont {Linehan}} \emph {et~al.},\ }\href
  {https://doi.org/10.1016/j.nima.2021.165955} {\bibfield  {journal} {\bibinfo
  {journal} {Nuclear Instruments and Methods in Physics Research Section A:
  Accelerators, Spectrometers, Detectors and Associated Equipment}\ }\textbf
  {\bibinfo {volume} {1031}},\ \bibinfo {pages} {165955} (\bibinfo {year}
  {2022})},\ \Eprint {https://arxiv.org/abs/2106.06622} {2106.06622}
  \BibitemShut {NoStop}%
\bibitem [{\citenamefont {Henriques}\ \emph {et~al.}(2019)\citenamefont
  {Henriques} \emph {et~al.}}]{NEXT2019_diffusion}%
  \BibitemOpen
  \bibfield  {author} {\bibinfo {author} {\bibfnamefont {C.~A.~O.}\
  \bibnamefont {Henriques}} \emph {et~al.} (\bibinfo {collaboration} {NEXT
  Collaboration}),\ }\href {https://doi.org/10.1007/JHEP01(2019)027} {\bibfield
   {journal} {\bibinfo  {journal} {Journal of High Energy Physics}\ }\textbf
  {\bibinfo {volume} {2019}},\ \bibinfo {pages} {27} (\bibinfo {year}
  {2019})}\BibitemShut {NoStop}%
\bibitem [{\citenamefont {Boehm}\ and\ \citenamefont
  {Vogel}(1992)}]{Boehm_Vogel_1992}%
  \BibitemOpen
  \bibfield  {author} {\bibinfo {author} {\bibfnamefont {F.}~\bibnamefont
  {Boehm}}\ and\ \bibinfo {author} {\bibfnamefont {P.}~\bibnamefont {Vogel}},\
  }\href@noop {} {\emph {\bibinfo {title} {Physics of Massive Neutrinos}}},\
  \bibinfo {edition} {2nd}\ ed.\ (\bibinfo  {publisher} {Cambridge University
  Press},\ \bibinfo {year} {1992})\BibitemShut {NoStop}%
\bibitem [{\citenamefont {Agostini}\ \emph {et~al.}(2020)\citenamefont
  {Agostini} \emph {et~al.}}]{Darwin2020_0vbb}%
  \BibitemOpen
  \bibfield  {author} {\bibinfo {author} {\bibfnamefont {F.}~\bibnamefont
  {Agostini}} \emph {et~al.} (\bibinfo {collaboration} {Darwin
  Collaboration}),\ }\href {https://doi.org/10.1140/epjc/s10052-020-8196-z}
  {\bibfield  {journal} {\bibinfo  {journal} {The European Physical Journal C}\
  }\textbf {\bibinfo {volume} {80}},\ \bibinfo {pages} {808} (\bibinfo {year}
  {2020})}\BibitemShut {NoStop}%
\bibitem [{\citenamefont {Ester}\ \emph {et~al.}(1996)\citenamefont {Ester},
  \citenamefont {Kriegel}, \citenamefont {Sander},\ and\ \citenamefont
  {Xu}}]{ester1996density}%
  \BibitemOpen
  \bibfield  {author} {\bibinfo {author} {\bibfnamefont {M.}~\bibnamefont
  {Ester}}, \bibinfo {author} {\bibfnamefont {H.-P.}\ \bibnamefont {Kriegel}},
  \bibinfo {author} {\bibfnamefont {J.}~\bibnamefont {Sander}},\ and\ \bibinfo
  {author} {\bibfnamefont {X.}~\bibnamefont {Xu}},\ }in\ \href@noop {} {\emph
  {\bibinfo {booktitle} {Proceedings of the Second International Conference on
  Knowledge Discovery and Data Mining}}},\ \bibinfo {series and number}
  {KDD'96}\ (\bibinfo  {publisher} {AAAI Press},\ \bibinfo {year} {1996})\ p.\
  \bibinfo {pages} {226–231}\BibitemShut {NoStop}%
\bibitem [{\citenamefont {Sander}\ \emph {et~al.}(1998)\citenamefont {Sander},
  \citenamefont {Ester}, \citenamefont {Kriegel},\ and\ \citenamefont
  {Xu}}]{sander1998density}%
  \BibitemOpen
  \bibfield  {author} {\bibinfo {author} {\bibfnamefont {J.}~\bibnamefont
  {Sander}}, \bibinfo {author} {\bibfnamefont {M.}~\bibnamefont {Ester}},
  \bibinfo {author} {\bibfnamefont {H.-P.}\ \bibnamefont {Kriegel}},\ and\
  \bibinfo {author} {\bibfnamefont {X.}~\bibnamefont {Xu}},\ }\href@noop {}
  {\bibfield  {journal} {\bibinfo  {journal} {Data mining and knowledge
  discovery}\ }\textbf {\bibinfo {volume} {2}},\ \bibinfo {pages} {169}
  (\bibinfo {year} {1998})}\BibitemShut {NoStop}%
\bibitem [{\citenamefont {Gallina}\ \emph {et~al.}(2022)\citenamefont
  {Gallina}, \citenamefont {Guan}, \citenamefont {Retiere} \emph
  {et~al.}}]{nEXO2022_SiPM}%
  \BibitemOpen
  \bibfield  {author} {\bibinfo {author} {\bibfnamefont {G.}~\bibnamefont
  {Gallina}}, \bibinfo {author} {\bibfnamefont {Y.}~\bibnamefont {Guan}},
  \bibinfo {author} {\bibfnamefont {F.}~\bibnamefont {Retiere}}, \emph
  {et~al.},\ }\href {https://doi.org/10.1140/epjc/s10052-022-11072-8}
  {\bibfield  {journal} {\bibinfo  {journal} {The European Physical Journal C}\
  }\textbf {\bibinfo {volume} {82}},\ \bibinfo {pages} {1125} (\bibinfo {year}
  {2022})}\BibitemShut {NoStop}%
\end{thebibliography}%

\end{document}